\documentclass[prb,twocolumn,showpacs,nofootinbib]{revtex4}
\usepackage[dvips]{graphicx}
\usepackage{color,array,dcolumn}
\usepackage{amsmath}
\usepackage{amssymb}

\begin{document}

\title{Adsorption of rare-gas atoms on Cu(111) and Pb(111) surfaces by
van der Waals-corrected Density Functional Theory}

\author{Pier Luigi Silvestrelli, Alberto Ambrosetti, 
        Sonja Grubisi\^{c},\cite{present} and Francesco Ancilotto} 
\affiliation{Dipartimento di Fisica, Universit\`a di Padova, via Marzolo 8, I--35131, 
Padova, Italy,
and DEMOCRITOS National Simulation Center, Trieste, Italy}

\begin{abstract}
\date{\today}
The DFT/vdW-WF method, recently developed to include the Van der
Waals interactions in Density Functional Theory (DFT) using
the Maximally Localized Wannier functions, is applied to the 
study of the adsorption of rare-gas atoms
(Ne, Ar, Kr, and Xe) on the Cu(111) and
Pb(111) surfaces, at three high-symmetry sites. We evaluate the
equilibrium binding energies and distances, and the induced 
work-function changes and dipole moments.
We find that, for Ne, Ar, and Kr on the Cu(111) surface the different
adsorption configurations are characterized by very similar binding energies,
while the favored adsorption site for Xe on Cu(111) is on top of a Cu atom, 
in agreement with previous theoretical calculations
and experimental findings, and in common with other close-packed
metal surfaces. Instead, the favored site is always the hollow one
on the Pb(111) surface, which therefore represents an interesting
system where the investigation of high-coordination sites is possible.
Moreover, the Pb(111) substrate is subject, upon rare-gas adsorption,
to a significantly smaller change in the work function
(and to a correspondingly smaller induced dipole moment) than Cu(111). 
The role of the chosen reference DFT functional and of different
Van der Waals corrections, and their dependence on different
rare-gas adatoms, are also discussed.
\end{abstract}

\maketitle

\section{Introduction}
Understanding adsorption processes on solid surfaces is essential to
design and optimize countless material applications, and
to interpret, for instance, scattering experiments and atomic-force microscopy.
In particular, the adsorption of rare-gas (RG) atoms on metal
surfaces is prototypical\cite{Bruch} for physisorption processes.
Basically, the weak binding of physisorbed closed electron-shell atoms,
such as RG atoms,
is due to an equilibrium between attractive, long-range van der Waals (vdW) 
interactions and short-range Pauli repulsion
acting between the electronic charge densities of the substrate 
and the adatoms.\cite{Vidali}

Up to now RG adsorption on many close-packed metal surfaces, such
as Ag(111), Al(111), Cu(111), Pd(111), Pt(111),.. have been 
extensively studied both
experimentally\cite{Gottlieb,Seyller,Narloch,Diehl} and 
theoretically,\cite{Diehl,Silva,DaSilva05,DaSilva,Betancourt,Lazic,Righi,Sun} 
while, to our knowledge, Pb has not, but for the experimental
measurements of Ferralis {\it et al.}\cite{Ferralis} and, very recently, 
the theoretical investigation of
Zhang {\it et al.},\cite{Zhang} who studied the tribological properties
of Ne and Kr on the Pb(111) surface. 
The Pb surface is important for practical applications: 
for instance, there
is considerable interest in the frictional (tribological) 
properties of gases on Pb at low temperatures; in particular, Pb 
is used\cite{Ferralis,Zhang,Bruschi} as a material for the electrodes 
and as adsorption surfaces in nanofriction experiments
because it is easy to grow a very uniform
film already at room temperature and to remove the surface contaminants
deposited over time on the electrodes,
thanks to its large diffusion coefficient.
The Pb(111) surface also exhibits interesting and unusual properties: 
for instance,
one striking finding is the drastic difference between the sliding 
friction of Ne and Kr mono- or multilayers.\cite{Bruschi,Zhang}

In principle, due to the non-directional character of the vdW
interactions that should be the dominant one in physisorption processes, 
surface sites that maximize the coordination of the RG adsorbate atom 
were expected to be the preferred ones,
so that it was usually assumed that
the adsorbate occupies the maximally coordinated {\it hollow} site.
This assumption was also based on the expectation
that the atom in the {\it hollow} site would be closer to the surface, thus
experiencing a more attractive potential; behind this 
is the notion that the repulsive potential at the
surface is proportional to the atomic charge density and the
natural assumption is that the charge density is highest at the
locations of the atoms, thus making the {\it top} site energetically unfavored.
Calculations where the total adatom-substrate interaction is described
by the sum of empirical binary potentials, which are widely used and
often give reasonable results for adsorption energies, seem to confirm this
expectation since the highly coordinated {\it hollow} sites naturally emerge as
the preferred adsorption sites for the adatoms.
However this picture has been questioned by many 
experimental\cite{Gottlieb,Seyller,Narloch} 
and theoretical\cite{DaSilva05,DaSilva,Betancourt,Lazic} 
recent studies which indicate that the actual scenario is 
more complex: in particular, for Xe and Kr a general tendency is 
found\cite{Diehl,DaSilva05,DaSilva,Betancourt,Lazic} for
adsorption on metallic surfaces in the low-coordination {\it top} sites
(this behavior was attributed\cite{Diehl,Bagus} to the delocalization
of charge density that increases the Pauli repulsion effect at the
{\it hollow} sites relative to the {\it top} site and lifts the potential well
upwards both in energy and height); for Ar the situation seems
to be less clear:\cite{DaSilva} for instance,
comparison of theoretical and experimental results\cite{Diehl}
would suggest that the {\it hollow} sites is still favored for Ar on Ag(111).

The importance of polarization effects to determine the favored
adsorption sites was pointed out by Da Silva {\it et al.},\cite{DaSilva} who
studied the interaction of RG adatoms with the Pd(111) surface:
in fact, for instance, for Xe the polarization is larger in the
on-top site, i.e. the larger induced dipole moment 
increases the attractive interaction between Xe and the
metal surface. 
Therefore, the dominant mechanisms 
appear to be polarization-induced attraction and site-dependent 
Pauli repulsion. The latter, being
weaker for the on-top site, stabilizes on-top adsorption.\cite{Silva}  

In spite of this recent substantial progress, the understanding of the
interaction of RGs with metal surfaces is not complete yet.\cite{Diehl}
It is not clear, for instance, whether a system exists where high-coordinated
site are always preferred.
Moreover, there have been relatively few studies of adsorption geometries for
the smaller RGs, although these are probably better candidates
for the observation of high-coordination sites, due to their reduced 
polarizability with respect to that of Xe or Kr: 
in fact, the considerable 
mismatch between the lattice constants of the smaller RGs and that of 
most metal surfaces 
cause most commensurate structures to have multiple atoms per
unit cell, so that the characterization and interpretation of such systems
is quite complex.

Density Functional Theory (DFT) is a well-established
computational approach to study
the structural and electronic properties of
condensed matter systems from first principles, and, in particular, to
elucidate complex surface processes such
as adsorptions, catalytic reactions, and diffusive motions. 
Although current density functionals are able to describe quantitatively
condensed matter systems at much lower computational cost than other
first principles methods, they fail\cite{Kohn} to properly describe
dispersion interactions. Dispersion forces originate from 
correlated charge oscillations in
separate fragments of matter and the most important 
component is represented by the $R^{-6}$ vdW interaction,\cite{london}
originating from correlated instantaneous dipole
fluctuations, which plays a fundamental role
in adsorption processes of fragments weakly interacting with a substrate
("physisorbed").

This is clearly the case for the present systems which can be divided 
into well separated fragments (RG atoms and the metal substrate)
with negligible electron-density overlap.
The local or semilocal character 
of the most commonly employed exchange-correlation functionals 
makes DFT methods unable to correctly predict binding energies and 
equilibrium distances within both the local density (LDA) and the 
generalized gradient (GGA) approximations.\cite{Riley}
As a consequence, the basic results often depend, even at a qualitative
level, on the adopted DFT functional:
for instance, in their ab initio study of the interaction of
RG adatoms with the Pd(111) surface, Da Silva {\it et al.}\cite{DaSilva}
found that the on-top
site preference is obtained by the LDA for all RG adatoms, while
the GGA functionals (in the PBE and PW91 schemes) yield the on-top site
preference for Xe, Kr, and He adatoms, but the {\it hollow} site for Ne and Ar.
Typically, in many physisorbed systems GGAs give only a shallow and 
flat adsorption well at large atom-substrate separations, while 
the LDA binding energy turns out to be not far from the experimental 
adsorption energy; however, since it is well known that LDA 
tends to overestimate the binding in systems with 
inhomogeneous electron density (and to underestimate the equilibrium
distances), the reasonable 
performances of LDA must be considered as accidental.
Therefore, a theoretical approach beyond the
DFT-LDA/GGA framework, that is able to properly describe vdW effects 
is required to provide more quantitative results.\cite{DaSilva}

In the last few years a variety of practical methods have been proposed
to make DFT calculations able to accurately describe vdW effects (for a
recent review, see, for instance, ref. \onlinecite{Riley}).
We have investigated by such a method
the adsorption of RG atoms on the Cu(111) and Pb(111) surfaces.
Cu(111) has been chosen because of the many 
experimental and theoretical data available
(especially for Xe-Cu(111)), which can
be compared with ours in such a way to validate the present approach;
as mentioned above, the less studied Pb(111) surface could be interesting
because, given the
relatively large Pb lattice constant (and hence nearest-neighbor
surface Pb-Pb distance) it represents a good candidate for a system 
where RG atoms are preferably adsorbed on 
{\it hollow} sites 
(the lattice constant of Pb is 4.95 \AA, compared to 4.09 \AA\ for
Ag, 4.05 \AA\ for Al, 3.92 \AA\ for Pt, 3.89 \AA\ for Pd, and 3.61 \AA\ for Cu).

\section{Method}
In this study we include vdW effects within a standard DFT approach by using
the method proposed in refs. \onlinecite{silvprl,silvsurf,silvmetodo}
(where further details can be found), hereafter referred to as
DFT/vdW-WF, by introducing an additional term in the exchange-correlation 
functional as originally proposed by Andersson {\it et al.}\cite{andersson} 
to describe the interactions between separate fragments.
This contribution, which effectively accounts for the
dispersion forces both in the uniform electron gas and separate atom 
limits, has the form :
\begin{equation}
E_{vdW}=-\sum_{n,l}f_{nl}(r_{nl})\frac{C_{6nl}}{r_{nl}^6}
\label{Evdw}
\end{equation}
with (in a.u.)
\begin{equation}
C_{6nl}=\frac{3}{16\pi^{3/2}}\int_{|\mathbf{r'}|<r'_c}d\mathbf{r'}
\int_{|\mathbf{r}|<r_c}d\mathbf{r} 
\frac{\sqrt{\rho_n(r)\rho_l(r')}}{\sqrt{\rho_n(r)}+\sqrt{\rho_l(r')}}\,.
\label{vdw}
\end{equation}
In the above formulas $r_{nl}$ is the distance between the two
separate fragments $n$ and $l$,
and $\rho_n(r)$ is the $n$-th fragment electronic density. 
The cutoff $r_c$ is introduced to remove the divergence 
of the integral, taking into account
that, at small momentum values, the interaction is highly 
damped.\cite{andersson}

In our approach all the fragment densities are conveniently 
rewritten in terms of the Maximally Localized
Wannier Functions (MLWFs), \{$w_n(r)$\}, i.e. $\rho_n(r)=w^2_n(r)$. 
The MLWFs can be obtained from the occupied
Kohn-Sham orbitals, generated by a standard DFT calculation, 
by means of a unitary transformation which minimizes
the functional\cite{wannier}
\begin{equation}
\Omega=\sum_n S^2_n=
\sum_n \left( <w_n|r^2|w_n>-<w_n|\mathbf{r}|w_n>^2\right).
\end{equation}
The unitary transformation conserves the total density, 
which is however partitioned into single
localized fragments, each of them being characterized by its spread $S_n$ 
and center of mass position $r_n$.
It is therefore possible to express the vdW correction 
(see eqs. \eqref{Evdw} and \eqref{vdw}) as a
sum of single contributions coming from each pair of Wannier functions 
belonging to different fragments,
by approximating the shape of the $n$-th Wannier function\cite{silvmetodo} 
with an H-like exponential. 

The DFT/vdW-WF method has been
already successfully applied to several systems, including 
small molecules, bulk, and
surfaces;\cite{silvprl,silvsurf,silvmetodo,silvhb,silvinter,ambrosetti}
in particular it allowed us to study the interaction of
Ar with graphite and
of Ar, He, and H$_2$ with Al surfaces,\cite{silvsurf,silvmetodo}
of water with the
Cl- and H-terminated Si(111) surfaces,\cite{silvinter} and of
RG atoms and water with graphite and graphene.\cite{ambrosetti}

We here apply the DFT/vdW-WF method to the case of adsorption of Ne, Ar, Kr, and
Xe atoms on the Cu(111) and Pb(111) surfaces.
All calculations have been performed 
with the Quantum-ESPRESSO\cite{ESPRESSO} ab initio package
(MLWFs have been generated as a post-processing calculation using
the WanT package\cite{WanT}). Similarly to DaSilva {\it et al.},\cite{DaSilva}
we modeled the clean and RG-covered metal surfaces 
using a periodically-repeated hexagonal supercell, 
with a $(\sqrt{3}\times \sqrt{3})R30^{\circ}$ structure and a surface slab
made of 15 metal (Cu or Pb) atoms distributed over 5 layers
(repeated slabs were
separated along the direction orthogonal to the surface by a vacuum region
of about 24 \AA). The Brillouin Zone has been sampled using a 
$6\times6\times1$ $k$-point mesh.
In this model system 
the RG coverage is 1/3, i.e. one RG adatom for each 3
metal atoms in the topmost surface layer. 
The $(\sqrt{3}\times \sqrt{3})R30^{\circ}$
structure has been indeed observed at low temperature by LEED for
the case of Xe adsorption on Cu(111) and Pd(111)\cite{Seyller} (actually, 
this is the simplest commensurate structure for RG monolayers on 
close-packed metal surfaces and the only one for which good 
experimental data exist), and it was adopted in most of 
the previous ab initio 
studies\cite{Silva,DaSilva05,DaSilva,Righi,Lazic,Zhang}.
Since the lateral interactions between RG adatoms do not play a critical role
in the RG adsorption site preference,\cite{DaSilva05} for the sake
of simplicity, we have used the
same structure also for the other RGs (Ne, Ar, and Kr) and 
in the case of adsorption on Pb(111) as well.  

The Pb or Cu surface atoms
were kept frozen (of course after a preliminary relaxation of the outermost
layers of the clean metal surfaces) and only the 
vertical coordinates of the RG atoms,
perpendicular to the surface, were optimized, this procedure being justified
by the fact that only minor surface atom displacements are observed
upon physisorption.\cite{DaSilva05,Zhang,Abad} 
Moreover, the RG atoms
were adsorbed on both sides of the slab: in this way the surface dipole
generated by adsorption on the upper surface of the slab is cancelled by the
dipole appearing on the lower surface, thus greatly reducing the
spurious dipole-dipole interactions between the periodically repeated images 
(previous DFT-based calculations have shown
that these choices are appropriate\cite{DaSilva,Sun}).
Note that, apparently, in their recent study of Ne and Kr on Pb(111),
Zhang {\it et al.}\cite{Zhang} have instead considered 
adsorption on a single side of the metal slab;
the effect of such a choice is non-negligible:
in fact, for instance, in the case of Xe on Pb(111), we find that the
(absolute value of the) binding energy is reduced by 7 meV 
(about 4 \%) with respect
to that obtained when Xe is adsorbed on both sides of the slab.
The results of ref. \onlinecite{Zhang} are thus likely affected by 
the artificial dipole-dipole interactions discussed above.

We have carried out calculations for various separations 
of the RG atoms adsorbed on high-symmetry sites, namely
{\it hollow} (on the center of the triangle formed by the 3 surface metal 
atoms contained in the supercell), {\it top} 
(on the top of a metal atom), and {\it bridge} (intermediate between
two nearest-neighbor metal atoms).
Actually, two kinds of {\it hollow} sites are present: HCP {\it hollows}, characterized
by having atoms directly beneath them in the next layer of atoms, and
FCC {\it hollows} where this condition does not apply; however
the HCP-{\it hollow} and the FCC-{\it hollow} sites can be considered equivalent 
for adsorption because of the small differences in the adsorption 
properties (for instance, Righi and Ferrario,\cite{Righi} using LDA, found
a difference of less than 1 meV in the adsorption energy and of 0.01 \AA\ in the
equilibrium distance for RGs adsorbed on Cu(111)). 
For a better accuracy, as done in previous applications on adsorption
processes,\cite{silvsurf,silvmetodo,silvinter,ambrosetti} we have also
included the interactions of the MLWFs of the physisorbed fragments not 
only with the MLWFs of the underlying surface, within the reference supercell,
but also with a sufficient
number of periodically-repeated surface MLWFs (in any case, given the
$R^{-6}$ decay of the vdW interactions, the convergence with
the number of repeated images is rapidly achieved).
Electron-ion interactions were described using norm-conserving
pseudopotentials: in the case of Pb and Cu we have explicitly included
14 and 11 valence electrons per atom, respectively (those coming from the
$5d^{10}$, $6s^2$, $6p^2$ atomic orbitals for Pb, and $3d^{10}$, 
$4s^1$ for Cu). 
As a reference DFT functional we chose PW91\cite{PW91} because it is
widely used in ab initio DFT calculations of solids and surfaces
and, in particular, was adopted in some previous simulations\cite{Lazic} of
Xe interacting with the Cu(111) surface, which facilitates 
comparison with the results of the present calculations
(note that typically PW91 gives similar results to that obtained 
by PBE,\cite{pbe} which represents another popular GGA functional).
Using the PW91 functional in test calculations with bulk Pb and Cu,
for the equilibrium properties the agreement with 
experimental estimates is comparable to that found in previous 
DFT calculations.\cite{DaSilva05,Lazic,Zhang}

By generating the MLWFs for the Cu(111) and Pb(111)
substrates, we observe a clear quantitative
separation between the spreads of the MLWFs describing $d$-like orbitals
and those of the (much more delocalized) MLWFs describing the $s$- and
$p$-like orbitals; moreover, given the high valence-electron
density, screening effects are certainly relevant in these metal surfaces.
Therefore, at variance
with previous calculations,\cite{silvsurf,silvmetodo,silvinter,ambrosetti} 
we have applied the DFT/vdW-WF correction by explicitly considering only the 
more localized MLWFs corresponding to the $d$-like orbitals, while
the $s$- and $p$-like electrons are supposed to give a 
screening-effect\cite{screening} contribution, which is taken into 
account by multiplying 
the vdW correction (the C$_6$ coefficients) by a simple Thomas-Fermi factor:
$ f_{TF} = e^{-2(z-z_s)/r_{_{TF}}}$
where $r_{_{TF}}$ is the Thomas-Fermi screening length relative 
to the electronic density of a uniform electron gas (''jellium model'')
equal to the average density of the $s$- and $p$-like electrons of the
present systems, $z_s$ is the average vertical position of the topmost Cu or 
Pb atoms, and $z$ is the vertical position, measured with respect to $z_s$,
of the adatom.
In practice it turns out that only the topmost metal layer 
gives a relevant contribution, while the effects of the
other ones is dramatically reduced by the exponential factor, in
line with the common expectation about screening effects in metal
surfaces.\cite{screening} This observation can be exploited to 
considerably reduce the computational cost of the vdW correction
since only the topmost MLWFs must be really taken into account.

\section{Results and Discussion}
In Tables I-VI results are reported for all the systems 
under consideration, for adsorption on {\it hollow}, {\it top}, and
{\it bridge} sites.
The {\it binding energy}, $E_b$, is defined as
\begin{equation}
E_b=1/2(E_{tot}-(E_s+2E_{RG}))
\end{equation}
where $E_{s,RG}$ represent the energies of the 
isolated fragments (the substrate and the RG atoms)
and $E_{tot}$ is the energy of the interacting system, including the 
vdW-correction term (the factors 2 and 1/2 are due to the adsorption of
RG atoms on both sides of the slab).

One should point out that
the experimentally measured {\it adsorption energy}, $E_a$,
includes not only the interaction of RG atoms with the substrate but
also lateral, vdW, RG-RG interactions;\cite{Sun} however in most of
previous calculations the mostly attractive lateral interaction 
contribution was not considered.
As pointed out, for instance, by Lee {\it et al.},\cite{Lee} 
who studied $n$-butane on 
transition-metal surfaces (another typical weak physisorption
system where vdW interaction is the only attractive force between the 
nonpolar molecule and the substrate) 
lateral adatom-adatom interaction energies can be as large as 25\% of
the total adsorption energy at full coverage.
$E_a$ is here defined as:

$E_a = E_b + (E_l - E_f) $, 

where $E_l$ is the total energy (per atom) of the 2D RG lattice
(that is as in the adsorption configurations but without the substrate)
and $E_f$ is the energy of an isolated RG atom. 
Clearly the quantity in parenthesis in the above formula represents
the lateral adatom-adatom interaction energy (per atom).
Note that, in their DFT study of Ne and Kr on Pb(111), 
Zhang {\it et al.}\cite{Zhang} seem instead to identify $E_a$ with $E_b$.

$E_b$ has been evaluated for several adsorbate-substrate distances; then
the equilibrium distances and the corresponding binding energies 
have been obtained by fitting the calculated points with
the function: $A\,e^{-Bz}-C_3/(z-z_0)^3$, $A$, $B$, $C_3$, and $z_0$ being 
adjustable parameters
(as illustrated for the Xe-Cu(111) and Xe-Pb(111) cases in Figs. 1 and 2).
Typical uncertainties in the fit are of the order of $0.05$ \AA$ $ 
for the distances and a few meV for the minimum binding energies.
Our results are compared to available theoretical and experimental estimates 
and to corresponding data obtained using a pure
PW91 functional, the simple LDA functional, and the ''seamless'' 
vdW-DF method 
of Langreth {\it et al.}\cite{Dion} (note that the vdW-DF method
has been also used in the recent DFT study of Zhang {\it et al.}\cite{Zhang}).
As can be seen in Figs. 1 and 2, and in Tables I and II, 
the effect of the vdW correction computed by DFT/vdW-WF is 
a much stronger bonding than with a pure PW91 scheme, 
with the formation of a clear minimum in the
binding energy curve at a shorter equilibrium distance.
In spite of the clear shortcomings of the pure PW91 scheme, in general
the preferred adsorption site seems to be correctly determined by the
latter, although
the differences between the binding energies of the
different adsorption sites are very small.

We have also computed $E_a$ (assuming a full monolayer coverage of RGs) 
in the case of Xe on Cu(111), 
where a RG overlayer in the $(\sqrt{3}\times \sqrt{3})R30^{\circ}$
structure is experimentally found\cite{Seyller} and in the
case of Xe on Pb(111), where the formation of a commensurate Xe monolayer
was also observed.\cite{Ferralis} 
As can be seen in Table III, all the methods, but pure PW91,
correctly predict a smaller $E_a$ 
on Pb(111) than on Cu(111), although the quantitative 
results considerably depend on the adopted scheme: in fact, pure PW91
clearly underestimates $E_a$, DFT/vdW-WF and vdW-DF give comparable results,
while LDA is close to DFT/vdW-WF and vdW-DF for Xe on Cu(111) but
underestimates for Xe on Pb(111): this can be explained by the fact that
LDA is not able to describe properly the lateral interactions of Xe adatoms
which are further from each other on Pb(111) than on Cu(111). 

Concerning the adsorption on the Cu(111) surface (see Table I), 
all the methods used 
predict that the {\it top} configuration is energetically favored 
in the case of Xe, while for Ne, Ar, and Kr the differences among
the binding energies of the different adsorption configurations
are quite small (using vdW-DF the same is true also for Xe);
since these differences are probably
comparable to the expected accuracy of the calculations,
a precise assignment of the favored adsorption site is not possible.  
In contrast, the {\it hollow} configuration is instead clearly
favored by all the methods (see Table II) in the case of the adsorption 
on Pb(111) of 
all the considered RG atoms (actually, with DFT/vdW-WF, for Ar on Pb(111) the 
{\it bridge} site appears to be lower in energy: however, given the
small difference with the energy of the {\it hollow} site, this
result should not be overemphasized).
Our results for Pb(111) are in qualitative agreement with those
of Zhang {\it et al.}\cite{Zhang} who predict that Ne and Kr indeed prefer
high-coordination {\it hollow} sites. 
Note that the energy difference between the {\it hollow} and {\it top} sites 
increases by subsequently considering the PW91, vdW-DF, DFT/vdW-WF, 
and LDA methods (see also Da Silva {\it et al.}\cite{DaSilva05}).

Interestingly, in the case where several experimental reference values
are available, namely Xe on Cu(111), our DFT/vdW-WF method 
performs better (considering both the binding and adsorption energy,
and the equilibrium distance, see Tables I, III, and IV) 
than all the other schemes: in fact LDA gives reasonable binding
energies but underestimates the equilibrium distances, while
vdW-DF underestimates the binding energies and overestimates
the equilibrium distances, in line with the behavior reported for systems
including a metallic surface.\cite{Vanin}  
Also note that, at a variance with the experimental
findings, the vdW-DF method predicts that the 
{\it top} site (see Table I) is only marginally favored (and the distance 
only marginally different) than the {\it hollow} ones; in general, 
for all the RG atoms on Cu(111) vdW-DF gives almost identical
binding energies for the {\it top} and {\it hollow} adsorption sites. 
In the case of RGs on Pb(111) the {\it hollow} structure is favored
also by vdW-DF, although the difference in the binding energy 
with respect to the {\it top} site is smaller than with the present
DFT/vdW-WF scheme
(the difference was instead larger, see the last column of Table II, 
in the study of Zhang 
{\it et al.}\cite{Zhang}, who used vdW-DF but with 
a reference DFT functional differing from ours by the exchange term).
In the case of Ar on Cu(111) and on Pb(111), we observe that 
our computed binding energies compare 
favorably with the estimates obtained, using a simple
Lennard-Jones potential, by Cheng {\it et al.},\cite{Cheng} 
who predicted a binding energy between 70 and 85 meV for
Ar on noble metals.

As expected, we find that, both for adsorption on Cu(111) and Pb(111), 
the binding energy increases by going from 
Ne to Xe, in line with the increasing polarizability of this atom sequence.
In particular, for several close-packed transition-metal surfaces 
the binding energy of Xe is found\cite{DaSilva} to be 
about 2 to 3 times larger than that of
Kr, and Ar, respectively, a behavior which is well reproduced by
our DF/vdW-WF method (the factors are 1.5 and 3, and 1.6 and 2.5, for
adsorption on Cu(111) and on Pb(111), respectively). 
This general behavior is also in line with the results of
Zhang {\it et al.}\cite{Zhang} 

Our energetic results are not far 
from the "best estimate" reported by Vidali {\it et al.} \cite{Vidali}
for Xe on Cu(111), i.e. a binding energy of -183 $\pm 10$ meV at
a distance of 3.60 $\pm 0.08$\AA\  
(these values represent averages over different 
theoretical/experimental estimates).
In their tables Vidali {\it et al.}\cite{Vidali} also report
for Ar on Cu(111) a binding energy of -85 meV at
a distance of 3.53 \AA\ and for Kr on Cu(111) a binding energy of 
-119 meV, in fair agreement with our results.
Lazic {\it et al.}\cite{Lazic} studied the adsorption of Xe on Cu(111) by
a DFT approach where vdW corrections were included using the
method of Andersson {\it et al.},\cite{andersson} 
using PW91 and PBE 
as reference DFT functionals (see the last column in Tables I and IV).
As can be seen, our results are much closer to the experimental estimate
than those of Lazic {\it et al.},\cite{Lazic} which tend to overestimate the
binding energy and underestimate the equilibrium distance.
The Xe-adsorbed Cu(111) surface has been also recently investigated by
Sun and Yamauchi\cite{Sun} using DFT with semiempirical vdW corrections:
they found reasonable equilibrium distances, however the computed
binding energy was very overestimated (it was even larger than that obtained
by LDA) and the favored adsorption site was incorrectly predicted to be
the {\it hollow} site, probably due to the use of semiempirical pair potentials
which favor close-packed structures and high coordinated sites (see
discussion above).

From Tables I, III, and IV, on can also see that the binding energies
are reasonably reproduced by the LDA scheme for RGs on Cu(111),
a behavior common to several physisorption systems.
However, as already outlined above,
this agreement should be considered accidental: the well-known
LDA overbinding, due to the overestimate of the long-range part of the
exchange contribution, somehow mimics the missing vdW interactions;
the equilibrium distances predicted by LDA are 
clearly underestimated since LDA
cannot reproduce the $R^{-6}$ behavior in the interaction potential.
For RGs on Pb(111), the LDA binding energies are instead 
underestimated as a consequence (as discussed above)
of the larger equilibrium distances than for RGs on Cu(111).

As already found elsewhere,\cite{DaSilva05,DaSilva} for all the used schemes,
the binding energies
correlate with the RG-metal distance: in fact, for a given RG,
the configurations having the strongest binding are characterized
by the shortest RG-substrate distance. 
Moreover, all the
methods predict that Ar and Xe adatoms get closer to the Cu(111) 
surface when adsorbed on {\it top} site, as found in some 
previous studies.\cite{DaSilva05,DaSilva,Righi} 
Remarkably, this behavior cannot be reproduced\cite{Diehl,DaSilva}
using a hard-sphere model, indicating that there is a significant
interaction between the Ar and Xe atoms and the Cu(111) surface
so that a simple stacking (hard-sphere) model of weakly or
noninteracting spheres is not valid (for comparison, in Tables IV and V we
also list the sums of the RG atoms and metal atom vdW literature radii).
Instead, for adsorption on Pb(111), the adatoms in the
{\it hollow} site are closer to the surface than in the {\it top} one,
in line with the usual behavior. These results can be easily 
elucidated by analyzing the parameters of the adopted fitting function
(see above), $A\,e^{-Bz}-C_3/(z-z_0)^3$ : we find that,
as a general rule, at the equilibrium distance, the repulsive potential 
term is weaker on the favored adsorption site (for instance the
{\it top} site for Xe on Cu(111) and the {\it hollow} one for Xe on Pb(111)),
in agreement with the results of Da Silva {\it et al.}\cite{DaSilva05} 

Ferralis {\it et al.}\cite{Ferralis} studied the structural and 
thermal properties of Xe on the Pb(111) surface by LEED.
They observed the formation of a Xe monolayer with 
an incommensurate hexagonal structure with
a lattice parameter similar to that found in bulk Xe (4.33 \AA);
this structure is aligned with the substrate lattice but has a larger
unit cell, similarly to the case of Xe on Ag(111), which is also an
aligned incommensurate monolayer.
They also found that the
heat of adsorption for the first Xe layer is -191 $\pm 10$ meV with an,
overlayer-substrate spacing of 3.95 $\pm 0.10$ \AA. 
Looking at Table III we found that our computed $E_a$ (-205.5 meV
for the {\it hollow} adsorption site) is close to the experimental value and
in better agreement than with the other methods,
although our model structure is not exactly the same observed 
experimentally; moreover, also the Xe-Pb(111) distance (3.93 \AA) is 
in excellent agreement 
(see Table V) with that estimated by Ferralis {\it et al.},\cite{Ferralis}
which gives further support to the reliability of our DF/vdW-WF method.
As expected, it has been found\cite{Ferralis} that
a hard-sphere model is unable to give a good description
of adsorption of Xe on Pb(111).
For Xe-Pb(111) the heat of adsorption is lower than for Xe on any 
surface measured so far,\cite{Ferralis} with the possible 
exception of Al(110) and
for alkali metals; a low heat of adsorption is not particularly 
surprising since the Pb atoms are much larger than most other
metals (the vdW radius of Pb is 2.02 \AA, compared to 1.72 \AA\ for
Ag, 1.72 \AA\ for Pt, 1.63 \AA\ for Pd, and 1.40 \AA\ for Cu),
implying that the repulsive Xe-Pb interaction
prevents the Xe from approaching the deeper part of the attractive
holding potential.
It must be noted that Ferralis {\it et al.}\cite{Ferralis} were unable
to determine the preferred adsorption site, the lack of 
satellite intensities in the LEED patterns indicating that
the overlayer is quite uniform and the corrugation is small.

An important quantity which often provides revealing details of 
the bonding mechanism in
adsorption processes is represented by the
electron density difference, 
$\Delta n({\bf r})= n_{RG/s}({\bf r})-n_s({\bf r})-n_{RG}({\bf r})$,
obtained from the electron density (at the equilibrium geometry)
of the RG on the substrate, of the clean substrate, and the isolated
RG monolayer, respectively. 
Our approach in this respect is not fully self-consistent because we use
the electron density obtained at a pure PW91 level, that is without
vdW corrections, however, the effects due to the
lack of self-consistency are expected to be negligible
because the rather weak and diffuse vdW interactions should not
substantially change the electronic charge distribution.\cite{Langreth07}
Plots of $\Delta n({\bf r})$ for Xe on Cu(111)
and Xe on Pb(111), both in the {\it hollow} and {\it top} site
(see Figs. 3 and 4), show that, in agreement with what found
previously\cite{DaSilva} for RGs on Pd(111),
the electron density redistribution is stronger on the Cu atoms
for the Xe on the {\it top} site than for the {\it hollow}; both sites
exhibit a significant depletion of electron density centered about the
Xe atom together with a slight density accumulation close to the center of 
the Xe atom, this effect being attributed\cite{DaSilva} to orthogonalization
of Xe states to the states of the substrate atoms.
Moreover, for Xe in the on-top site, there is a significant electron density
accumulation between the Xe atom and the topmost surface layer.
Interestingly, there is a clear tendency of 
Xe to induce a much larger charge delocalization on the
Cu(111) surface than on Pb(111), in line with the 
delocalization mechanism invoked\cite{Diehl,Bagus} to explain
the preference for the {\it top} adsorption site on Cu(111).

Since polarization effects are assumed to play a key role 
in determining the favored adsorption sites,\cite{DaSilva,Righi}
we have also computed the change of the work function, $\Delta W$, of the
Cu(111) and Pb(111) substrate upon adsorption of RG atoms.
The work functions have been calculated as the difference
between the averaged electrostatic Coulomb potential at the midpoint of 
the vacuum region of the slab and the Fermi energy:
\cite{Binggeli} 
for the clean Cu(111) and Pb(111) surfaces we estimate a work function
of 4.85 and 3.86 eV, respectively, in excellent agreement with the
reference values, that are in the range from 4.90 to 5.01 eV\cite{Fall} 
for Cu(111), and 3.83 eV for Pb(111).\cite{Sun08} 
$\Delta W$ can be related to the dipole moment induced in the
substrate by the presence of the RG adatom, $\Delta \mu$,
using the Helmholtz equation:\cite{Schmidt}

\begin{equation}
\Delta \mu = {1 \over {12\pi}} {{A_{(1\times 1)}}\over {\Theta}} 
\Delta W\;,
\label{deltamu}
\end{equation}

where $ A_{(1\times 1)}$ is the area of the $(1\times 1)$
surface unti cell (in \AA$^2$) and $\Theta$ is the RG coverage;
if $\Delta W$ is given in eV, then $\Delta \mu$
is in debyes. In our case $\Theta = 1/3$, so that 
$\Delta \mu = \sqrt{3} a_0^2/16\pi \Delta W$, where $a_0$ is the
Cu or Pb lattice constant.
Our computed $\Delta W$ and $\Delta \mu$ values are listed in Table VI.
In agreement with previous ab initio calculations,\cite{DaSilva,DaSilva05} we 
find that the RG adsorption induces a decrease in the work function,
thus indicating that the RG atoms behave as adsorbates with an
effective positive charge; note that this is consistent with 
the depletion of the electron density about the Xe atom discussed above,
which corresponds to an induced surface dipole moment that points out
of the surface.
For Xe on Cu(111) our estimated $\Delta W$ and $\Delta \mu$ values
(see Table VI) agree well with the experimental 
estimates\cite{Zeppenfeld} of -0.60 eV and -0.24 D, respectively. 
As can be seen in Table VI, the absolute value of $\Delta \mu$
increases from Ne to Xe, because the corresponding electronic
polarizabilities increase, and is larger for the optimal adsorption
site, for instance the {\it top} for Xe on Cu(111) and the {\it hollow} 
for Xe on Pb(111).
Moreover, it is considerably larger on Cu(111) than on Pb(111)
in line with the energetic analysis reported above, that 
indicated a stronger interaction of RGs with the Cu(111) surface 
than with Pb(111).

Zhang {\it et al.}\cite{Zhang} explain the much larger mobility of 
Ne overlayers on Pb(111), as observed in friction experiments, than of Kr 
overlayers on the basis of the different activation energies 
which characterize the lateral motion of Ne and Kr atoms on the Pb(111)
surface.
The activation energies for a monolayer can be directly calculated
from the difference in the binding energy of the adatom between the 
favored ({\it hollow}) site and the transition state, which is expected to 
correspond to the {\it bridge} site.\cite{Zhang}
Considering the differences between the binding energy
of the {\it hollow} and {\it bridge} configurations for Ne and Kr on Pb(111), 
we qualitatively confirm the trend observed by Zhang {\it et al.},\cite{Zhang}
being our estimated activation energies (1.3 meV for Ne and 6.0 meV for Kr)
of the same order of magnitude as those 
reported in ref. \onlinecite{Zhang} (0.7 meV for Ne and 2.5 meV for Kr).
However, such small energy values are comparable to
(or even smaller than) the expected
accuracy of the computed binding energies, thus making quantitative
estimates of the hopping probabilities\cite{Zhang} 
(which depend exponentially on the aforementioned activation energies) 
rather questionable.

\begin{figure}
\centerline{
\includegraphics[scale=0.35, angle=-90]{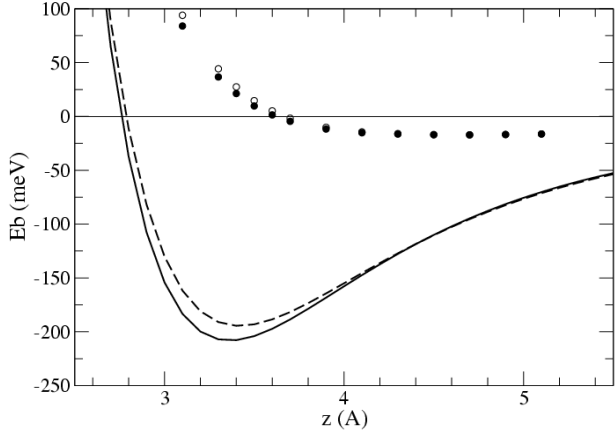}
}
\caption{Binding energy of Xe on Cu(111) in the {\it top} and
{\it hollow} configuration using pure PW91 
(full and empty circles, respectively) 
and DFT/vdW-WF (solid and dashed line, respectively), as a function of the
distance $z$ from the surface.}
\label{fig1}
\huge
\end{figure}
                      
\begin{figure}
\centerline{
\includegraphics[scale=0.35, angle=-90]{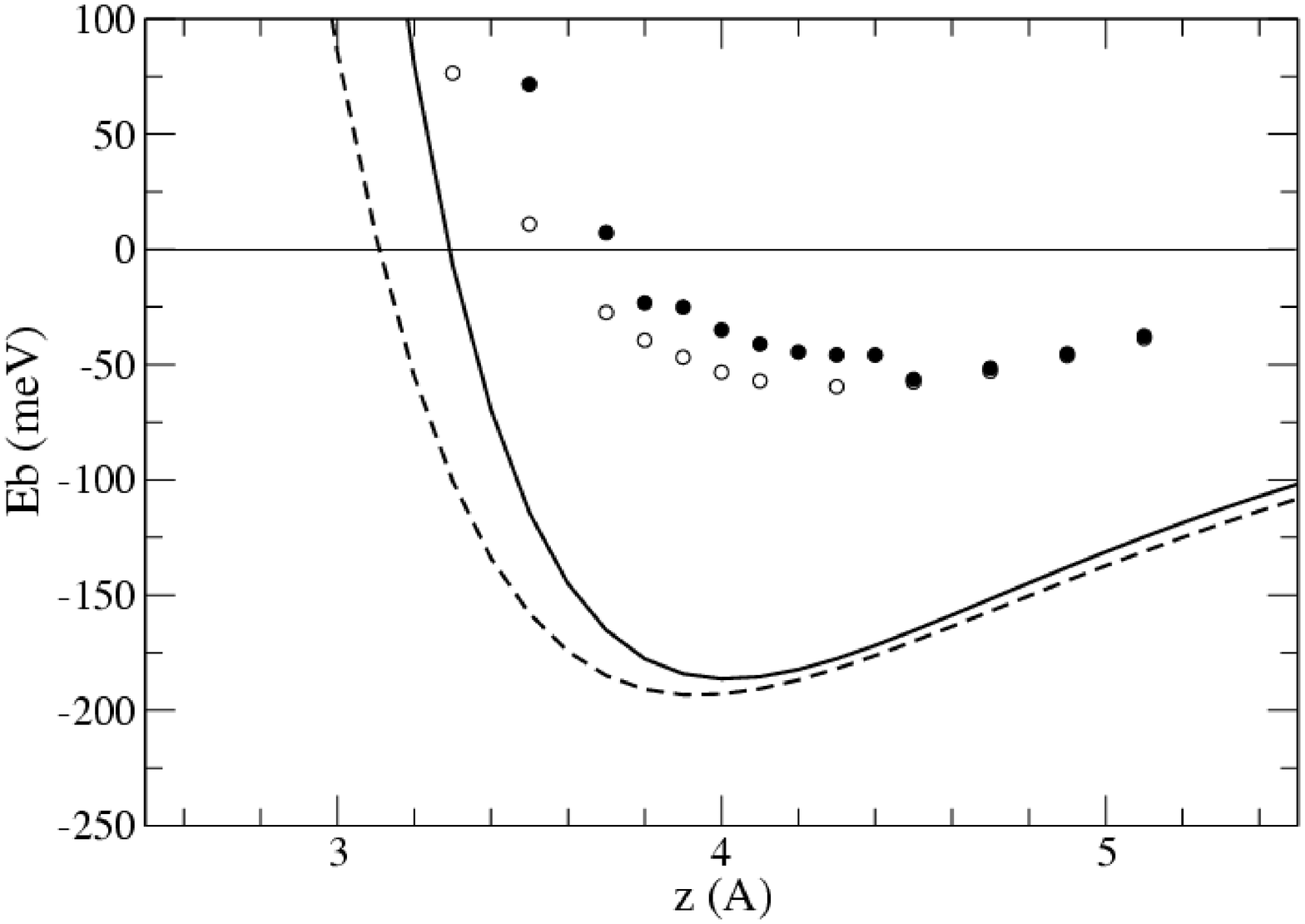}
}
\caption{Binding energy of Xe on Pb(111) in the {\it top} and
{\it hollow} configuration using pure PW91 
(full and empty circles, respectively) 
and DFT/vdW-WF (solid and dashed line, respectively), as a function of the
distance $z$ from the surface.}
\label{fig2}
\huge
\end{figure}
                      
\begin{figure}
\centerline{
\includegraphics[scale=0.4]{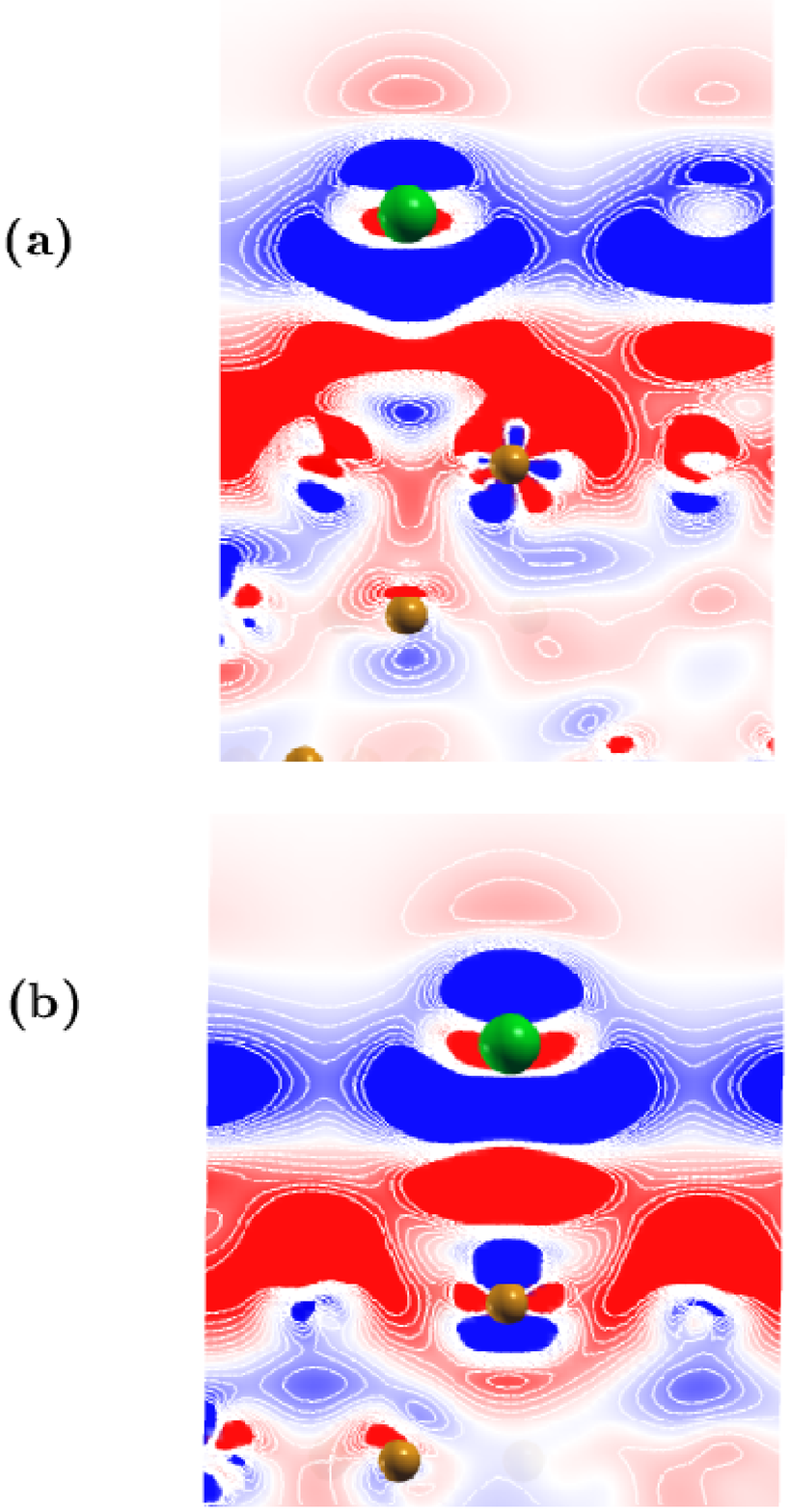}
}
{\vskip -0.8cm}
\caption{Electron density difference of Xe on Cu(111) 
in (a) {\it hollow} and (b) {\it top} site 
shown in a plane  perpendicular to the surface, within
the range of $\pm 1 \times 10^{-4} e/{\rm bohr}^3$.
Red (light grey) and blue (dark grey) represent electron accumulation
and depletion, respectively. The green and orange spheres indicate
the Xe and Cu atoms, respectively.}
\label{fig3}
\huge
\end{figure}
                      
\begin{figure}
\centerline{
\includegraphics[scale=0.4]{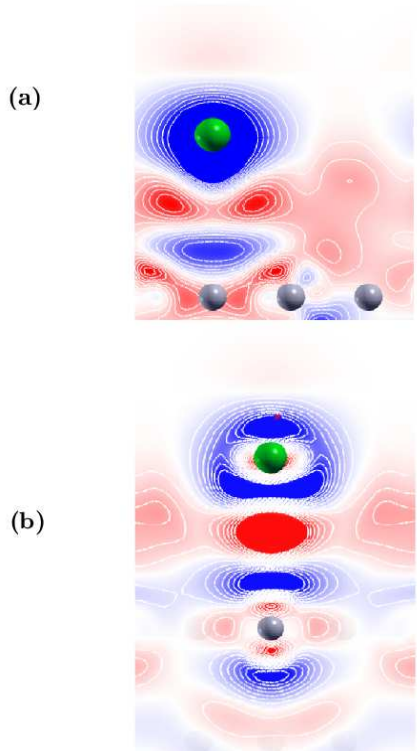}
}
{\vskip -0.8cm}
\caption{Electron density difference of Xe on Pb(111) 
in (a) {\it hollow} and (b) {\it top} site 
shown in a plane perpendicular to the surface, within
the range of $\pm 1 \times 10^{-4} e/{\rm bohr}^3$.
Red (light grey) and blue (dark grey) represent electron accumulation
and depletion, respectively. The green and grey spheres indicate
the Xe and Pb atoms, respectively.}
\label{fig4}
\huge
\end{figure}

\section{Conclusions}
In summary, by analyzing the results of our study of the adsorption of RG
atoms on the Cu(111) and Pb(111) surfaces,
one can conclude that the inclusion of the
vdW corrections by the DFT/vdW-WF method systematically improves upon
the estimates for the binding energies as obtained by
a standard GGA approach. In particular, using a pure PW91 functional
the binding is underestimated in all cases, while 
equilibrium distances are overestimated.
For all the system considered the vdW correction term represents the
dominant part of the binding energy, although, particularly for RG adsorption
on Pb(111), the pure PW91 approach gives a substantial contribution.
However, vdW interactions appear not to play a critical role in the 
adsorption site preference (the same result has been obtained by
Zhang {\it et al.}\cite{Zhang} studying the interaction of 
Ne and Kr on Pb(111)): 
Xe on Cu(111) clearly prefers the {\it top} site, while for Ne, Ar,
an Kr on Cu(111) the differences in binding energies relative
to different adsorption sites are so small that is not easy
to attribute a definitive preference; instead, 
the {\it hollow} configuration tends to be preferred
for adsorption of all the considered RGs on Pb(111), in agreement 
with previous calculations and experimental observations.\cite{Ferralis,Zhang} 
Moreover, the Pb(111) substrate is subject, upon rare-gas adsorption,
to a significantly smaller change in the work function,
and to a correspondingly smaller (in absolute value) induced dipole moment, 
than Cu(111). 
Given these relevant peculiarities of the Pb(111) surface, where 
the {\it hollow} site is undoubtedly favored for adsorption of RG atoms, 
this surface would represent an ideal substrate to study,
both theoretically and experimentally, high-coordination 
adsorption sites.

\section{Acknowledgements}
We thank very much F. Costanzo and F. Toigo for useful discussions.

\vfill
\eject

\begin{table}
\caption{Binding energy, $E_b$ in meV, of RG atoms on the Cu(111) 
surface computed using the standard DFT-PW91
calculation, and including the vdW corrections using our DFT/vdW-WF method,
compared to the LDA result, 
the vdW-DF method by Langreth {\it et al.}\cite{Dion} 
and available theoretical and
experimental (in parenthesis) reference data.}
\begin{center}
\begin{tabular}{|l|c|c|c|c|c|}
\hline
system & PW91 & DFT/vdW-WF & LDA & vdW-DF & ref. \\ \tableline
\hline
Ne-Cu(111) {\it hollow}  &-17.6&-31.6 & -55.7& -56.1 &  --- \\
Ne-Cu(111) {\it top}     &-17.5&-31.1 & -55.4& -55.9 &  --- \\
Ne-Cu(111) {\it bridge}  &-17.6&-31.0 & -55.3& -56.1 &  --- \\
\hline
Ar-Cu(111) {\it hollow}  &-13.0&-67.8 & -88.9&-106.6 &  --- \\
Ar-Cu(111) {\it top}     &-13.0&-71.9 & -94.5&-106.3 &  -85$^a$\\
Ar-Cu(111) {\it bridge}  &-13.0&-70.6 & -89.4&-106.4 &  --- \\
\hline
Kr-Cu(111) {\it hollow}  & -20.3&-134.2&-117.6& -135.7&  --- \\
Kr-Cu(111) {\it top}     & -20.3&-131.1&-126.0& -135.8& -119$^a$\\
Kr-Cu(111) {\it bridge}  & -20.3&-130.0&-118.4& -135.7&  --- \\
\hline
Xe-Cu(111) {\it hollow}  &-22.9&-194.5&-199.3& -167.4& -276$^b$, -268$^c$\\
Xe-Cu(111) {\it top}     &-23.1&-208.1&-221.9& -167.7& -280$^b$, -183$^a$, -277$^c$ (-190$^c$)\\
Xe-Cu(111) {\it bridge}  &-17.1&-191.2&-201.0& -167.4& -278$^b$\\
\hline
\end{tabular}
\tablenotetext[1]{ref.\onlinecite{Vidali}.}
\tablenotetext[2]{ref.\onlinecite{Lazic}.}
\tablenotetext[3]{ref.\onlinecite{Silva}.}
\end{center}
\label{table1}
\end{table}
\vfill
\eject

\begin{table}
\caption{Binding energy, $E_b$ in meV, of RG atoms on the Pb(111) 
surface computed using the standard DFT-PW91
calculation, and including the vdW corrections using our DFT/vdW-WF method,
compared to the LDA result, the vdW-DF method by 
Langreth {\it et al.}\cite{Dion} 
and available theoretical and experimental (in parenthesis) reference data.}
\begin{center}
\begin{tabular}{|l|c|c|c|c|c|}
\hline
system & PW91 & DFT/vdW-WF & LDA & vdW-DF & ref. \\ \tableline
\hline
Ne-Pb(111) {\it hollow}  &-31.2&-59.8& -49.4& -71.4& -51.6$^a$ \\
Ne-Pb(111) {\it top}     &-27.8&-49.1& -42.9& -63.3& -46.8$^a$ \\
Ne-Pb(111) {\it bridge}  &-19.8&-58.5& -49.1& -64.6&   --- \\
\hline
Ar-Pb(111) {\it hollow}  &-23.5&-82.4& -78.3&-100.8&   --- \\
Ar-Pb(111) {\it top}     &-22.1&-75.0& -64.2& -95.3&   --- \\
Ar-Pb(111) {\it bridge}  &-22.7&-84.5& -76.6&-100.1&    --- \\
\hline
Kr-Pb(111) {\it hollow}  &-30.8&-132.8& -98.8&-136.9&-134.9$^a$ \\
Kr-Pb(111) {\it top}     &-29.1&-109.8& -81.6&-130.9&-125.1$^a$ \\
Kr-Pb(111) {\it bridge}  &-24.0&-126.8& -96.7&-136.1&     --- \\
\hline
Xe-Pb(111) {\it hollow}  &-59.6&-193.5&-142.0&-192.2&-172.6$^a$ \\
Xe-Pb(111) {\it top}     &-56.3&-186.4&-116.1&-186.4&  --- \\
Xe-Pb(111) {\it bridge}  &-52.7&-188.9&-138.6&-191.2&    --- \\
\hline
\end{tabular}
\tablenotetext[1]{ref.\onlinecite{Zhang}.}
\tablenotetext[2]{ref.\onlinecite{Ferralis}.}
\end{center}
\label{table2}
\end{table}
\vfill
\eject

\begin{table}
\caption{Adsorption energy ($E_a$, see text for the definition), in meV, 
of Xe atoms on the Cu(111) and Pb(111) 
surfaces computed using the standard DFT-PW91
calculation, and including the vdW corrections using our DFT/vdW-WF method,
compared to the LDA result, the vdW-DF method by 
Langreth {\it et al.}\cite{Dion} and available 
experimental (in parenthesis) reference data.}
\begin{center}
\begin{tabular}{|l|c|c|c|c|c|}
\hline
system & PW91 & DFT/vdW-WF & LDA & vdW-DF & ref. \\ \tableline
\hline
Xe-Cu(111) {\it hollow}  &-51.4&-289.3&-297.3& -268.9&  --- \\
Xe-Cu(111) {\it top}     &-51.6&-302.9&-319.9& -269.2&  (-227$^a$)\\
Xe-Cu(111) {\it bridge}  &-45.6&-286.0&-299.0& -268.9& --- \\
\hline
Xe-Pb(111) {\it hollow}  &-62.5&-205.5&-147.9&-252.2& (-191$^a$)\\
Xe-Pb(111) {\it top}     &-59.2&-198.4&-122.0&-246.4&  --- \\
Xe-Pb(111) {\it bridge}  &-55.6&-200.9&-146.9&-251.2&  --- \\
\hline
\end{tabular}
\tablenotetext[1]{ref.\onlinecite{Ferralis}.}
\end{center}
\label{table3}
\end{table}
\vfill
\eject

\begin{table}
\caption{Equilibrium RG adatom-surface distance, in \AA, on the Cu(111) 
surface computed using the standard DFT-PW91
calculation, and including the vdW corrections using our DFT/vdW-WF method,
compared to the LDA result, the vdW-DF method 
by Langreth {\it et al.}\cite{Dion} and available theoretical and
experimental (in parenthesis) reference data; the sum, $s$, of the 
vdW radii of the RG atom and the Cu atom is also reported.}
\pagestyle{empty}
\begin{center}
\begin{tabular}{|l|c|c|c|c|c|c|}
\hline
system & PW91 & DFT/vdW-WF & LDA & vdW-DF &ref. & $s$ \\ \tableline
\hline
Ne-Cu(111) {\it hollow}     &3.90&3.59&3.10& 3.70 &   ---     & 2.94 \\
Ne-Cu(111) {\it top}        &3.90&3.57&3.09& 3.68 &   ---     & 2.94 \\
Ne-Cu(111) {\it bridge}     &3.90&3.60&3.10& 3.68 &   ---     & 2.94 \\
\hline
Ar-Cu(111) {\it hollow}     &4.50&3.48& 3.19& 3.90&   ---    & 3.28 \\
Ar-Cu(111) {\it top}        &4.50&3.45& 3.15& 3.86&   3.53$^a$& 3.28 \\
Ar-Cu(111) {\it bridge}     &4.50&3.43& 3.19& 3.86&   ---   & 3.28 \\
\hline
Kr-Cu(111) {\it hollow}     &4.50&3.32& 3.21& 3.99 &  ---    &3.42 \\
Kr-Cu(111) {\it top}        &4.50&3.36& 3.17& 3.99 &  ---    &3.42 \\
Kr-Cu(111) {\it bridge}     &4.50&3.35& 3.20& 3.99 &  ---    &3.42 \\
\hline
Xe-Cu(111) {\it hollow}     &4.70&3.42&3.00&4.10&  3.40$^b$, 3.31$^c$   &3.56 \\
Xe-Cu(111) {\it top}        &4.40&3.36&2.90&4.09&  3.45$^b$, 3.2$^d$, 3.25$^c$ (3.60$^e$)& 3.56 \\
Xe-Cu(111) {\it bridge}     &4.70&3.41&3.00&4.10&  ---   & 3.56 \\
\hline
\end{tabular}
\tablenotetext[1]{ref.\onlinecite{Vidali}.}
\tablenotetext[2]{ref.\onlinecite{Sun}.}
\tablenotetext[3]{ref.\onlinecite{Silva}.}
\tablenotetext[4]{ref.\onlinecite{Lazic}.}
\tablenotetext[5]{ref.\onlinecite{Seyller}.}
\end{center}
\label{table4}
\end{table}
\vfill
\eject

\begin{table}
\caption{Equilibrium RG adatom-surface distance, in \AA, on the Pb(111) 
surface computed using the standard DFT-PW91
calculation, and including the vdW corrections using our DFT/vdW-WF method,
compared to the LDA result, the vdW-DF method 
by Langreth {\it et al.}\cite{Dion} compared to the 
LDA result, and available theoretical and and available theoretical and
experimental (in parenthesis) reference data; the sum, $s$, of the 
vdW radii of the RG atom and the Pb atom is also reported.}
\pagestyle{empty}
\begin{center}
\begin{tabular}{|l|c|c|c|c|c|c|}
\hline
system & PW91 & DFT/vdW-WF & LDA & vdW-DF & ref. & $s$ \\ \tableline
\hline
Ne-Pb(111) {\it hollow}     &3.80&3.41&3.10&3.70& 3.5$^a$ & 3.56 \\
Ne-Pb(111) {\it top}        &4.00&3.68&3.40&3.90& 3.8$^a$ & 3.56 \\
Ne-Pb(111) {\it bridge}     &3.80&3.36&3.27&3.50& --- & 3.56 \\
\hline
Ar-Pb(111) {\it hollow}     &4.40&3.68&3.40&4.00& --- & 3.90 \\
Ar-Pb(111) {\it top}        &4.40&4.04&3.60&4.22& --- & 3.90 \\
Ar-Pb(111) {\it bridge}     &4.50&3.77&3.43&4.10& --- & 3.90 \\
\hline
Kr-Pb(111) {\it hollow}     &4.40&3.69&3.40&4.14& 3.8$^a$ & 4.04 \\
Kr-Pb(111) {\it top}        &4.40&3.98&3.70&4.24& 3.9$^a$ & 4.04 \\
Kr-Pb(111) {\it bridge}     &4.30&3.79&3.51&4.13& --- & 4.04 \\
\hline
Xe-Pb(111) {\it hollow}     &4.30&3.93&3.50&4.30&  (3.95$^b$) & 4.18 \\
Xe-Pb(111) {\it top}        &4.50&4.02&3.70&4.30& --- & 4.18 \\
Xe-Pb(111) {\it bridge}     &4.70&3.93&3.55&4.31& --- & 4.18 \\
\hline
\end{tabular}
\tablenotetext[1]{ref.\onlinecite{Zhang}.}
\tablenotetext[2]{ref.\onlinecite{Ferralis}.}
\end{center}
\label{table5}
\end{table}
\vfill
\eject

\begin{table}
\caption{Work-function change, in eV, and induced dipole moment 
(in parenthesis), in debyes, for RGs adatoms on the Cu(111) and
Pb(111) surfaces, at equilibrium geometries.}
\pagestyle{empty}
\begin{center}
\begin{tabular}{|c|c|c|}
\hline
system & {\it hollow} & {\it top}  \\ \tableline
\hline
Ne-Cu(111)                  &-0.04 (-0.02)&-0.03 (-0.01)\\
Ar-Cu(111)                  &-0.28 (-0.13)&-0.37 (-0.17)\\
Kr-Cu(111)                  &-0.54 (-0.24)&-0.37 (-0.17)\\
Xe-Cu(111)                  &-0.53 (-0.24)&-0.57 (-0.26)\\
\hline
Ne-Pb(111)                  &-0.03 (-0.03)&-0.03 (-0.03)\\
Ar-Pb(111)                  &-0.10 (-0.08)&-0.03 (-0.03)\\
Kr-Pb(111)                  &-0.11 (-0.09)&-0.05 (-0.04)\\
Xe-Pb(111)                  &-0.13 (-0.11)&-0.04 (-0.03)\\
\hline
\end{tabular}
\end{center}
\label{table6}
\end{table}
\vfill
\eject

\end{document}